# Brownian Particles and Matter Waves

Nicos Makris

Dept. of Civil and Environmental Engineering, Southern Methodist University, Dallas, Texas, 75275

**Abstract**

In view of the remarkable progress in microrheology to monitor the random motion of Brownian particles with size as small as few nanometers, in association that de Broglie matter waves have been experimentally observed for large molecules of comparable nanometer size; we examine whether Brownian particles can manifest a particle-wave duality without employing a priori arguments from quantum decoherence. First, we examine the case where Brownian particles are immersed in a memoryless viscous fluid with a time-independent diffusion coefficient; and the requirement for the Brownian particles to manifest a particle-wave duality leads to the untenable result that the diffusion coefficient has to be proportional to the inverse time; therefore, diverging at early times. This finding agrees with past conclusions by Grabert et al., Phys. Rev. A **119** (1979), that quantum mechanics is not equivalent to a Markovian diffusion process. Next, we examine the case where the Brownian particle is trapped in a harmonic potential well with and without dissipation. Both solutions of the Fokker-Plank equation for the case with dissipation, and of the Schrödinger equation for the case without dissipation lead to the same physically acceptable result—that for the Brownian particle to manifest a particle-wave duality, its mean kinetic energy $K_B T/2$ needs to be ½ the ground-state energy, $E_0 = \frac{1}{2}\hbar\omega$ of the quantum harmonic oscillator. Our one-dimensional calculations show that for this to happen, the trapping needs to be very strong so that a Brownian particle with mass $m$ and radius $R$ needs to be embedded in an extremely stiff solid with shear modulus, $G$ proportional to $(m/R)(K_B T/\hbar)^2$.

## INTRODUCTION

Small probe particles immersed in a fluid or solid material experience thermally driven Brownian motion that originates from the random collisions of the molecules of the surrounding material on the Brownian particles (microspheres). The diameter of these randomly moving Brownian particles can be as small as few nanometers ($10^{-9}$ m) [1-3]; while, their thermal fluctuations have been monitored with dynamic light scattering (DLS) and diffusing wave spectroscopy (DWS) [4-7]; or with laser interferometry [8-13] with a nanometer spatial resolution and sub-microsecond temporal resolution.

Given the continuous random fluctuations of the position and velocity of these Brownian particles from the collisions of the molecules of the surrounding material, their position, $x$ at any given time, $t$, is described by a probability density function, $p(x,t)$ as was first shown by Einstein [14] upon solving a one-dimensioanl diffusion equation which predicted the long term (diffusing regime) of Brownian particles immersed in a memoryless, Newtonian viscous fluid. Similarly, the position probability density function $p(x,t)$ of a Brownian particle trapped in a harmonic potential well with dissipation (damped harmonic oscillator) can be calculated by solving the more elaborate Fokker-Plank equation [15-19]. Alternatively, the evolution of the collective motion of Brownian particles immersed in any isotropic linear viscoelastic material can be evaluated for all time scales (including the early ballistic regime) by integrating the Langevin equation in terms of ensemble averages [4, 5, 7, 11-13, 15, 20-25].

The wave-like behavior of elementary particles was proposed by de Broglie [26] nearly a century ago, and his hypothesis was confirmed in 1927 with the seminal experiments by Thomson and Reid [27] and Davisson and Germer [28], showing that the electron is also a matter-wave. Following the 1927 electron diffraction experiments, superposition of de Broglie matter waves has been observed for particles way more massive than the electron. For instance, Arndt et al. [29] reported the observation of de Broglie wave interference of $C_{60}$ molecules with a diameter of approximately 0.7 nm (0.7x$10^{-9}$ m) and a mass that is 1.3 million (1.3x$10^6$) times larger than that of the electron. More recently, Shayeghi et al. [30] observed de Broglie wave interference from a natural antibiotic composed of 15 amino-acids (Gramicidin) with a mass larger than 2.6 times the mass of the $C_{60}$ molecule. Experiments in a Kapitza-Dirac-Tabot-Lau interferometer (KDTLI) have shown matter-wave interference with a functionalized tetraphenylporphyrin molecule that combines 810 atoms into one particle with a molecular weight exceeding 10,000 Da [31]—that is 14 times more massive than the $C_{60}$ molecule; while, more recently interference of de Broglie waves has been reported from matter-waves emitted from a molecular library with mass beyond 25,000 Da [32]. These remarkable experimental findings in matter-wave interferometry which confirm the wave-particle duality of complex organic molecules and inorganic clusters, that their size is in the nanometer scale [33, 34], suggest that perhaps small Brownian particles with comparable nanometer size and with inherent random behavior could manifest a particle-wave duality. Accordingly, in this paper we examine whether the position probability density, $p(x,t)$ of Brownian particles as computed from statistical mechanics with the Fokker-Planck equation [15-19] can also be the position probability density $\psi(x,t)\psi^*(x,t) = |\psi(x,t)|^2$, where $\psi(x,t)$ is a complex-valued wave function that satisfies the Schrödinger matter-wave equation [35-37].

The idea of relating non-relativistic quantum mechanics to a stochastic process goes back to the work of Nelson [38]. Nevertheless, Grabert et al. [39] showed that quantum mechanics is not equivalent to a Markovian diffusion process as advanced in [38]. The same conclusion is reached herein by showing that regardless how small the size of the Brownian particles is, when immersed in a memoryless, Newtonian viscous fluid, their position probability density, $p(x,t)$, cannot be associated with the position probability density $\psi(x,t)\psi^*(x,t) = |\psi(x,t)|^2$ of a matter-wave. The interactions of the energetic state of the Brownian particle with the molecules of the surrounding viscous fluid influence the statistics of its position since its mean-square displacement (variance) grows linearly with time—a manifestation of decoherence [40-43]. In contrast, our study shows that when a Brownian particle is embedded in an elastic solid, trapped in a harmonic potential well with or without dissipation, the requirement for the particle to manifest a particle-wave duality leads to the physically acceptable result that its mean kinetic energy $\frac{1}{2}K_BT$ needs to be ½ the energy of the ground state of the quantum harmonic oscillator, $E_0 = \frac{1}{2}\hbar\omega$. This result is reached by following two independent routes. First, we follow a top-down approach where the position probability density, $p(x,t)$ of a Brownian particle trapped in a harmonic potential with dissipation is described with the Fokker-Planck equation and the resulting Gaussian distribution function for $p(x,t)$ is set equal to the position probability density from a matter-wave $|\psi(x,t)|^2 = \psi_1^2(x,t) + \psi_2^2(x,t)$. Second, we follow a bottom-up approach by examining the solution of an oscillating quantum particle (wave pocket) trapped in a harmonic potential without dissipation [36, 37] and by demanding that the average kinetic energy of the oscillating wave-pocket to be equal to the average kinetic energy of a Brownian particle $= \frac{1}{2}K_BT$. Our calculations from both routes reach precisely the same results and show that for this to happen the trapping needs to be very strong, so that a Brownian particle with mass $m$ and radius $R$ needs to be embedded in an extremely stiff solid with shear modulus, $G$ proportional to $(m/R)(K_BT/\hbar)^2$. Cooling reduces the value of the shear modulus $G$ needed to offer the required trapping; nevertheless, in view of the various challenges associated with cooling schemes for interferometry with massive particles [33,34], our study concludes that the potential observation of de Broglie matter waves emitted from trapped Brownian nanoparticles remains a formidable task.

## AVERAGE KINETIC ENERGY FROM STATISTICAL AND QUANTUM MECHANICS

Regardless whether the Brownian particles are immersed in a memoryless viscous fluid or embedded in a viscoelastic solid (trapped in a damped harmonic potential well), the probability density, $p(x,t)$ of finding a Brownian particle at position *x* at time *t* is [14-19, 21-25]

$$p(x,t) = \frac{1}{\sigma(t)\sqrt{2\pi}} e^{-\frac{1}{2}\frac{(x-\mu(t))^2}{\sigma^2(t)}} \tag{1}$$

where $\mu(t)$ is the mean value and $\sigma^2(t)$ is the variance of the Gaussian distribution given by equation (1), and in the general case are time-dependent moments.

In the event that the Brownian particle with mass *m*, manifests a particle-wave duality, its spatial and temporal evolution should also be described with Schrödinger's matter-wave equation [34-36],

$$i\hbar \frac{\partial \psi(x,t)}{\partial t} = -\frac{\hbar^2}{2m}\frac{\partial^2 \psi(x,t)}{\partial x^2} + V(x)\psi(x,t) \tag{2}$$

where $\psi = \psi_1 + i\psi_2$ is the complex-valued wave function, $\hbar = h/2\pi$ with $h = 6.62607 \times 10^{-34}$ Joules s =Plank's constant and *V* is the potential energy of the particle. Upon multiplying equation (2) from the left with the complex conjugate of $\psi$ that is $\psi^* = \psi_1 - i\psi_2$ and integrate over the entire space,

$$\int_{-\infty}^{\infty} \psi^*\left(i\hbar \frac{\partial}{\partial t}\right)\psi dx = -\frac{\hbar^2}{2m}\int_{-\infty}^{\infty}\psi^*\frac{\partial^2 \psi}{\partial x^2}dx + \int_{-\infty}^{\infty}\psi^* V\psi dx \tag{3}$$

Recognizing that the left-hand side of equation (3) is the expectation value of the energy operator $\hat{E} = i\hbar \frac{\partial}{\partial t}$, and that the last integral is the expectation (mean) value of the potential *V*, equation (3) assumes the form

$$\langle \hat{E} \rangle = \frac{\hbar^2}{2m}\int_{-\infty}^{\infty}\left|\frac{d\psi}{dx}\right|^2 dx + \langle V \rangle \tag{4}$$

The integral in equation (4) derives from equation (3) upon integrating by parts $\int_{-\infty}^{\infty}\psi^*\frac{\partial^2 \psi}{\partial x^2}dx$, which is the average kinetic energy of the quantum/Brownian nanoparticle

$$\langle E_{KIN} \rangle = \frac{\hbar^2}{2m} \int_{-\infty}^{\infty} \left(\frac{d\psi}{dx}\right)\left(\frac{d\psi^*}{dx}\right) dx = \frac{\hbar^2}{2m} \int_{-\infty}^{\infty} \left|\frac{d\psi}{dx}\right|^2 dx \qquad (5)$$

For the Brownian particle to manifest a particle-wave duality, its position probability density $p(x,t)$ given by equation (1) should equal its position probability density $|\psi(x,t)|^2 = \psi(x,t)\psi^*(x,t)$ from its quantum mechanical description

$$p(x,t) = |\psi(x,t)|^2 = \psi(x,t)\psi^*(x,t) = \psi_1^2(x,t) + \psi_2^2(x,t) \qquad (6)$$

From equation (6), $\sqrt{p(x,t)}$ is the hepotenuse of the right-angle triangle shown in Figure 1

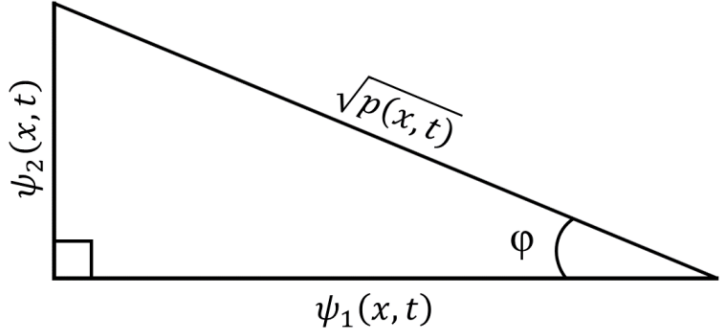

**Figure 1**. Relation of the probability densities, $p(x,t)$ from statistical mechanics and $|\psi(x,t)|^2 = \psi_1^2(x,t) + \psi_2^2(x,t)$ from a quantum mechanical description for a Brownian particle to be a matter-wave.

From the right-angle triangle shown in Figure 1

$$\psi_1(x,t) = \cos\varphi \sqrt{p(x,t)} = \frac{\cos\varphi}{\sqrt{\sigma(t)}(2\pi)^{1/4}} e^{-\frac{1}{4}\frac{(x-\mu(t))^2}{\sigma^2(t)}} \qquad (7a)$$

and

$$\psi_2(x,t) = \sin\varphi \sqrt{p(x,t)} = \frac{\sin\varphi}{\sqrt{\sigma(t)}(2\pi)^{1/4}} e^{-\frac{1}{4}\frac{(x-\mu(t))^2}{\sigma^2(t)}} \qquad (7b)$$

The space derivatives of the components of the complex wave function $\psi(x,t) = \psi_1(x,t) + i\psi_2(x,t)$ are

$$\frac{d\psi_1(x,t)}{dx} = -\frac{\cos\varphi}{\sqrt{\sigma(t)}\,(2\pi)^{\frac{1}{4}}}\frac{1}{2}\frac{(x-\mu(t))}{\sigma^2(t)}e^{-\frac{1}{4}\frac{(x-\mu(t))^2}{\sigma^2(t)}} \qquad (8a)$$

$$\frac{d\psi_2(x,t)}{dx} = -\frac{\sin\varphi}{\sqrt{\sigma(t)}\,(2\pi)^{\frac{1}{4}}}\frac{1}{2}\frac{(x-\mu(t))}{\sigma^2(t)}e^{-\frac{1}{4}\frac{(x-\mu(t))^2}{\sigma^2(t)}} \qquad (8b)$$

From equations (8a) and (8b), the amplitude, $\left|\frac{d\psi}{dx}\right|^2 = \left(\frac{d\psi_1}{dx}\right)^2 + \left(\frac{d\psi_2}{dx}\right)^2$, appearing in the integral of equation (5) is

$$\left|\frac{d\psi}{dx}\right|^2 = \left(\frac{d\psi_1}{dx}\right)^2 + \left(\frac{d\psi_2}{dx}\right)^2 = \frac{1}{4\sigma^4(t)}\frac{(x-\mu(t))^2}{\sigma(t)\sqrt{2\pi}}e^{-\frac{1}{2}\frac{(x-\mu(t))^2}{\sigma^2(t)}} \qquad (9)$$

Accordingly, the average kinetic energy of the quantum/Brownian particle expressed with equation (5) is

$$\langle E_{KIN}\rangle = \frac{\hbar^2}{2m}\int_{-\infty}^{\infty}\left|\frac{d\psi}{dx}\right|^2 dx = \frac{\hbar^2}{8m\sigma^4(t)}\int_{-\infty}^{\infty}\frac{(x-\mu(t))^2}{\sigma(t)\sqrt{2\pi}}e^{-\frac{1}{2}\frac{(x-\mu(t))^2}{\sigma^2(t)}} dx = \frac{\hbar^2}{8m\sigma^2(t)} \qquad (10)$$

since the integral in the right-hand side of equation (10) is merely the second moment of the Gaussian = $\sigma^2(t)$.

Given that the material within which the Brownian nanoparticle is immersed/embedded is in thermal equilibrium, from the equipartition theorem [15, 23-25], the average kinetic energy of the Brownian particle in one dimension is $\langle E_{KIN}\rangle = \frac{1}{2}K_B T$ where $K_B$ is Boltzmann's constant and $T$ is the equilibrium temperature of the material surrounding the Brownian-particles. Accordingly, from eq. (10),

$$\langle E_{KIN}\rangle = \frac{\hbar^2}{8m\sigma^2(t)} = \frac{1}{2}K_B T \qquad (11)$$

Equation (11) suggests that for a Brownian particle with mass, $m$ to manifest a particle-wave duality, the variance $\sigma^2$ of its position probability density given by equation (1) shall be a constant, independent of time.

$$\sigma^2 = \frac{\hbar^2}{4mK_BT} \tag{12}$$

In this way the statistics of its position probability $p(x,t)$, which encode the interactions of the energetic state of the Brownian particle with the surrounding material, are time-independent—which is shown to be a prerequisite for the randomly moving nanoparticle to maintain a quantum coherence.

## BROWNIAN PARTICLES IMMERSED IN A MEMORYLESS, NEWTONIAN VISCOUS FLUID

The random, thermally driven Brownian motion of a collection of M particles immersed in an isotropic fluid that is in thermal equilibrium can be described either by calculating the spatial and temporal evolution of the number-of-particles density, $\rho(x,t)$ (number of Brownian particles per unit volume), or by integrating, in terms of ensemble averages, the equation of motion of the Brownian particles when subjected to a randomly fluctuating force—known as the Langevin force [4, 5, 11-13, 20-25].

Einstein [14] followed the first approach, and formulated a one-dimensional diffusion equation for the number-of-particles density, $\rho(x,t)$, in which the diffusion coefficient, $D$ is a time-independent constant. The solution of the diffusion equation for a collection of $M$ Brownian particles that start from the origin ($x = 0$) at time, $t = 0$, leads to a Gaussian distribution for the number-of-particles density, $\rho(x,t)$; therefore, the probability density, $p(x,t)$ of finding a Brownian particle at position $x$ at time $t$ is $\rho(x,t)/M$.

$$p(x,t) = \frac{1}{\sqrt{4\pi Dt}} e^{-\frac{x^2}{4Dt}} \tag{13}$$

Equation (13), which is equation (1) for a zero mean ($\mu(t) = 0$), indicates that the time-dependent variance of the zero-mean, one-dimensional Gaussian distribution, $p(x,t)$ is $\sigma^2(t) = 2Dt$. More generally, the variance, $\sigma^2(t) = \langle r^2(t)\rangle - \langle r(t)\rangle^2$ of the zero-mean ($\langle r(t)\rangle = 0$) normal distribution given by Eq. (13), when x is replaced by $r$, is

$$\sigma^2 = \frac{1}{M}\sum_{j=1}^{M} r_j^2(t) = \langle r^2(t)\rangle = 2NDt \tag{14}$$

where $N \in \{1,2,3\}$ is the number of spatial dimensions and D is the time-independent diffusion coefficient. Upon replacing the expression of the variance $\sigma^2(t) = 2Dt$ given by equation (14) for $N=1$, into equation (11) we reach a diffusion coefficient proportional to the inverse time.

$$D = \frac{\hbar^2}{8mK_BT}\frac{1}{t} \tag{15}$$

The result of equation (15), where the diffusion coefficient $D = D(t)$ is proportional to the inverse time is a contradiction, given that in the position probability density, $p(x,t)$ offered by equation (13), that our analysis builds upon, the diffusion coefficient $D$ is a time-independent constant. Accordingly, Brownian particles when immersed in a Newtonian, viscous fluid cannot manifest a particle-wave duality. This result validates the predictions from decoherence theory [40-43]—that the random momentum kicks exerted on the nanoparticle "blur" its energetic state in the momentum representation; and is in agreement with the conclusion of Grabert et al. [39] who showed that quantum mechanics is not equivalent to a Markovian diffusion process.

Time-dependent diffusion coefficients with finite values at early times have been proposed to model anomalous diffusion [44], or with reference to equation (14), as the time derivative of the mean-square displacement [45-47]. In contrast, equation (15) suggests a diverging diffusion coefficient for a viscous fluid, a result that is physically untenable. On the other hand, equation (12) suggests that a necessary condition for a Brownian particle to manifest a particle-wave duality is for the variance of the position distribution of the Brownian particle $\sigma^2 = \hbar^2/(4mK_BT)$ to be a time-independent constant, a situation that eventually happens when the Brownian particle is trapped in a harmonic potential well.

**BROWNIAN PARTICLE TRAPPED IN A HARMONIC POTENTIAL WITH DISSIPATION**

We now consider the case when the Brownian particles are trapped in a harmonic potential with dissipation [11-13, 15, 21, 22, 48, 49]. In this case in addition to the viscous drag $F_d = -6\pi R\eta\frac{dx}{dt}$, the Brownian particle is subjected to an elastic trapping force $F_s = -kx = -m\omega^2 x$ [11-13, 21, 22, 48-51]. Given that the strong elastic trapping force $F_s$ and that the dissipative drag force, $F_d$ dominate over inertia forces, the position probability density function of the trapped Brownian particle is described by the Fokker-Planck equation [15-19]

$$\frac{\partial p(x,t)}{\partial t} = -\frac{\partial}{\partial x}[f(x)p(x,t)] + \frac{\partial^2}{\partial x^2}[D(x,t)p(x,t)] \qquad (16)$$

where $f(x)$ is the drift coefficient and $D(x,t) > 0$ is the diffusion coefficient [15-19]. For a Brownian particle trapped in a damped harmonic potential well,

$$f(x) = -kMx \text{ and } D(x,t) = D = K_B T M \qquad (17)$$

where $M = \frac{v}{F_d} = \frac{1}{6\pi R \eta}$ is the mobility of the Brownian particle and $k = 6\pi R G = m\omega^2$ is the stiffness of the harmonic oscillator in which $\eta$ is the shear viscosity and $G$ is the elastic shear modulus of the solid-like material within which the Brownian particle is embedded [1, 3, 7-9, 48-51]. Substitution of the expressions of the drift and diffusion coefficients given by equation (17) into equation (16) yields.

$$\frac{\partial p(x,t)}{\partial t} = kM\frac{\partial}{\partial x}[xp(x,t)] + K_B T M \frac{\partial^2 p(x,t)}{\partial x^2} \qquad (18)$$

The general solution of equation (18) is [15]

$$p(x,t) = \left[\frac{k}{2\pi K_B T(1-e^{-2kMt})}\right]^{1/2} exp\left[-\frac{1}{2}\frac{k(x-x_0 e^{-kMt})^2}{K_B T(1-e^{-2kMt})}\right] \qquad (19)$$

By setting $\sigma_0^2 = \frac{K_B T}{k} = \frac{K_B T}{(m\omega^2)}$, that is half the asymptotic mean-square displacement of the Brownian particle trapped in a damped harmonic potential [11-13, 21, 22, 48, 49], the position probability density function of the trapped Brownian particle given by equation (18) assumes the standard Gaussian form given by equation (1).

$$p(x,t) = \frac{1}{\sqrt{2\pi}}\frac{1}{\sigma^2(t)}e^{-\frac{1}{2}\frac{x_0^2}{\sigma^2(t)}\left(\frac{x}{x_0}-e^{-kMt}\right)^2} \qquad (20)$$

where

$$\sigma^2(t) = \sigma_0^2(1-e^{-2kMt}) = \frac{K_B T}{k}(1-e^{-2kMt}) \qquad (21)$$

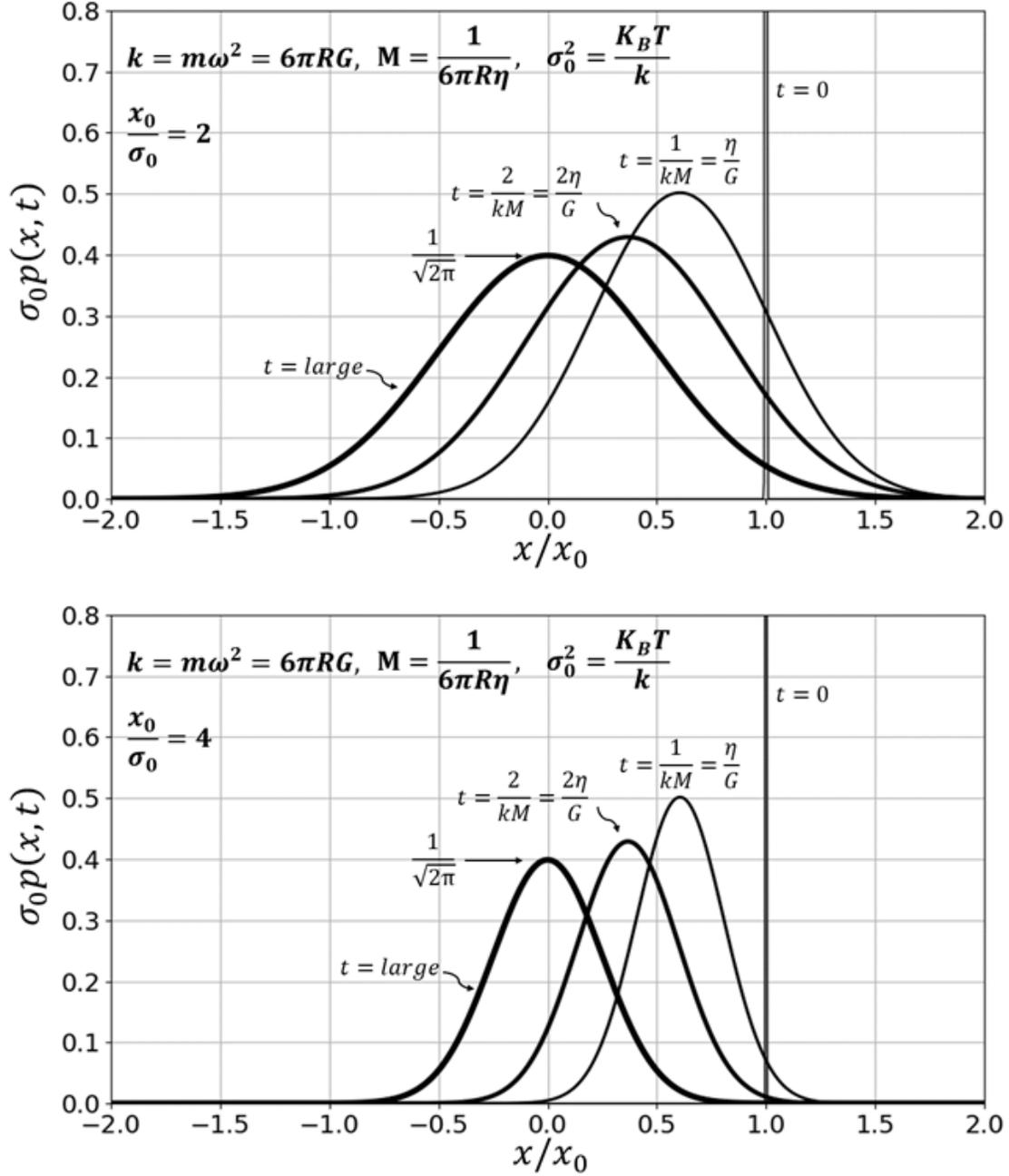

**Figure 2**. Evolution of the position probability density given by equation (20) (delocalization) of a Brownian particle trapped in a harmonic potential well with restoring force $F_s = -kx = -m\omega^2 x$, while experiencing a drag force $F_d = -6\pi R\eta \, dx/dt$. Top: $\frac{x_0}{\sigma_0} = 2$. Bottom: Wider well with $\frac{x_0}{\sigma_0} = 4$. At large times, the variance, $\sigma^2 = \sigma_0^2 = \frac{K_B T}{k} = \frac{K_B T}{6\pi RG}$ becomes time-independent which is a prerequisite for the randomly moving nanoparticle to maintain a stable coherence between separate space-time points.

is the time-dependent variance of the Gaussian distribution which eventually assumes the constant value $\sigma_0^2 = K_B T/k = K_B T/(m\omega^2)$. Similarly, the time-dependent mean value, $\mu(t) = x_0 e^{-kMt}$, eventually tends to zero. The inverse relaxation time, $kM$ appearing in the exponent of the exponentials in equations (19)-(21) is

$$kM = \frac{6\pi RG}{6\pi R\eta} = \frac{G}{\eta} \qquad (22)$$

and is merely the ratio of the macroscopic shear modulus $G$, to the macroscopic shear viscosity $\eta$ of the solid-like viscoelastic material within which the Brownian particles are embedded. Accordingly, after a short time the statistics (mean and variance) of the position are constant. Figure 2 plots the evolution of the position distribution density function, $p(x,t)$ given by equation (20) at early times, at times equal to one and two times the relaxation time $\frac{1}{kM} = \frac{\eta}{G}$, and at large times. Returning to equation (11), which equates the average kinetic energies as computed by quantum mechanic and statistical mechanics, the substitution of $\sigma^2(t)$ given by equation (21) gives the frequency of the harmonic oscillator that is required for the trapped Brownian nanoparticle to manifest a particle-wave duality

$$\omega = \frac{2K_B T}{\hbar}\sqrt{1 - e^{-2kMt}} = \frac{2K_B T}{\hbar}\sqrt{1 - e^{-2\frac{G}{\eta}t}} \qquad (23)$$

In the long term, $\omega = 2K_B T/\hbar$, leading to the result that the average kinetic energy of the Brownian particle, $\frac{1}{2}K_B T = \frac{1}{2}\left(\frac{\hbar\omega}{2}\right)$ – that is equal to ½ the energy of the ground state of the quantum harmonic oscillator, $E_0 = \frac{1}{2}\hbar\omega$ – a result that is in agreement with the equipartition theorem given that the remaining $\frac{1}{2}\left(\frac{\hbar\omega}{2}\right)$ will be equal to the average potential energy of the trapped Brownian nanoparticle.

For the limiting case of an undamped potential well, $(\eta = 0)$, the frequency of the Brownian/quantum nanoparticle assures immediately the value, $\omega = 2K_B T/\hbar$, – a result that is recovered in the next section by following a bottom-up route where we examine the solution of an oscillating quantum particle (wave pocket) trapped in a quantum harmonic potential.

The expected wave-length of the de Broglie matter-wave emitted from the trapped quantum/Brownian nanoparticle is calculated from the quadratic dispersion relation that results from the Hamiltonian of the quantum/Brownian nanoparticle [36,37], $\omega = \frac{\hbar}{2m}\left(\frac{2\pi}{\lambda}\right)^2$ in association with the long-term result of equation (23).

$$\lambda = \frac{\pi\hbar}{\sqrt{mK_BT}} = \frac{h}{2}\frac{1}{\sqrt{mK_BT}} \tag{24}$$

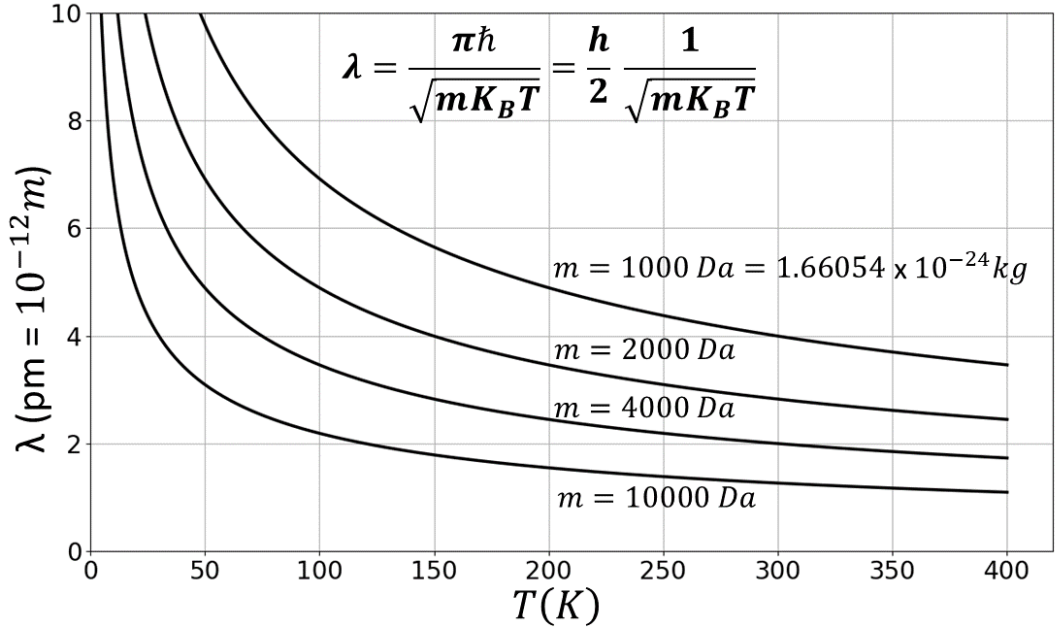

**Figure 3.** Theoretical values of the de Broglie wave-length, $\lambda$ of the matter-wave emitted from a Brownian particle with mass, $m$ trapped in a harmonic potential well which result upon equating the position probability densities $p(x,t)$ derived by solving the Fokker-Planck equation (18) and $|\psi(x,t)|^2 = \psi(x,t)\psi^*(x,t)$ where $\psi(x,t)$ is the wave function that satisfies the Schrödinger equation (2).

Figure 3 plots the necessary theoretical values of the de Broglie wave-length $\lambda$ offered by equation (24) so that a Brownian nanoparticle with mass $m$ when trapped in a harmonic potential well to manifest a particle-wave duality.

The long term expression of the frequency, $\omega = \frac{2K_BT}{\hbar}$ of the Brownian particle trapped in the harmonic potential with stiffness $k = 6\pi RG$ [1, 9, 48-51] allows for an estimate, within the context of the one-dimensional analysis presented herein, of the elastic shear modulus, $G$ of the viscoelastic solid that is needed to offer the required strong trapping for a Brownian particle with mass $m$ and radius $R$ to manifest a particle-wave duality. Accordingly,

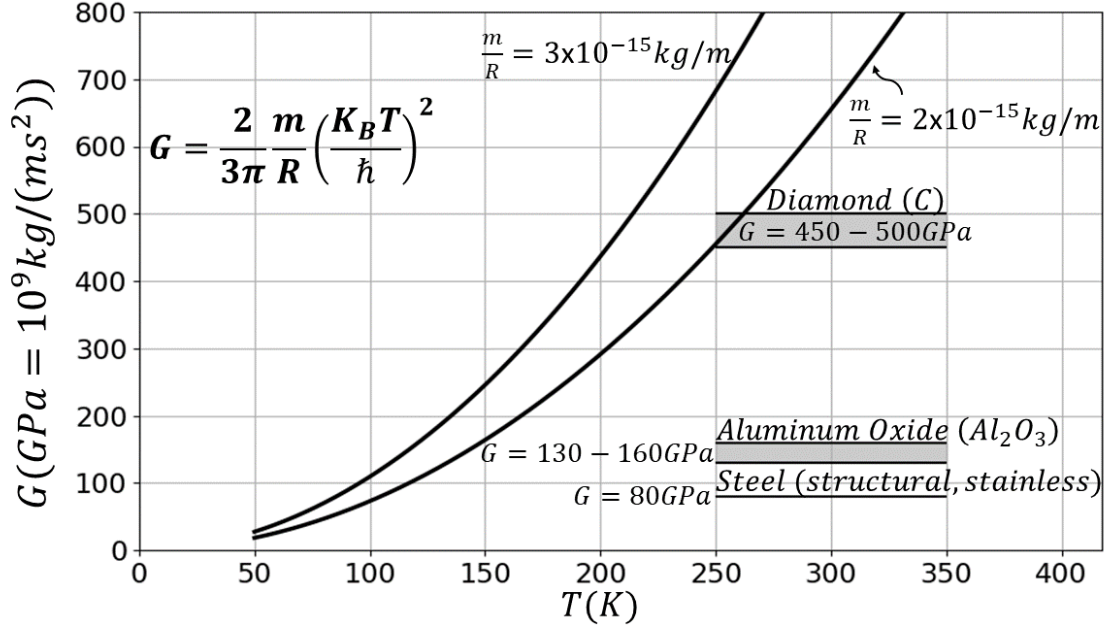

**Figure 4.** Values of the shear modulus, $G$ of a solid-like material that is needed to provide the strong trapping for Brownian particles with $m/R = 2x10^{-15}$ kg/m and $3x10^{-15}$ kg/m to manifest a particle-wave duality.

$$G = \frac{k}{6\pi R} = \frac{m\omega^2}{6\pi R} = \frac{2}{3\pi}\frac{m}{R}\left(\frac{K_B T}{\hbar}\right)^2 \quad (25)$$

Typical values for the ratio $m/R$ are available in the literature. For instances for the $C_{60}$ molecule $\frac{m}{R} \approx \frac{720 Da}{0.35 nm} = 3.42 \times 10^{-15}$ kg/m [29, 33, 34]; for the PFNS10 ($C_{60}[C_{12}F_{15}]_{10}$) molecule, $\frac{m}{R} \approx \frac{6910 Da}{1.7 nm} = 6.75 \times 10^{-15}$ kg/m, for the TPPF152 ($C_{168}H_{94}F_{152}O_8N_4S_4$) molecule, $\frac{m}{R} \approx \frac{5310 Da}{3.0 nm} = 2.94 \times 10^{-15}$ kg/m, whereas for a polypeptide Gramicidin A, $\frac{m}{R} \approx \frac{1860 Da}{1.5 nm} = 2.05 \times 10^{-15}$ kg/m [30, 52].

Figure 4 plots the values of the elastic shear modulus $G$ offered by equation (25), as result from the one-dimensional analysis presented herein, that is necessary for the embedded Brownian nanoparticle with $\frac{m}{R} = 2x10^{-15}$ and $3x10^{-15}$ kg/m to manifest a particle-wave duality. At room temperature the predicted values of the shear modulus, $G$ of the solid-like material that are needed to offer the required strong trapping are extremely high—of the order of ten times the shear modulus of steel which is in the range of 80 GPa. Cooling reduces the value of the shear modulus $G$ needed to offer the required trapping; while, it contributes to maintain the quantum coherence of the trapped nanoparticle. Nevertheless, cooling schemes for interferometry with massive particles are still in an early development stage [34, 53, 54], making the potential observation of de Broglie matter waves emitted from a trapped Brownian nanoparticle a formidable task.

**BROWNIAN PARTICLES EMBEDDED IN AN UNDAMPED ELASTIC SOLID**

For a Brownian nanoparticle with mass $m$ trapped in an undamped harmonic potential to also be a matter-wave, its evolution in space and time shall be described with the time-dependent Schrödinger equation given by equation (2), which for a harmonic potential, $V = \frac{1}{2}kx^2 = \frac{1}{2}m\omega^2 x^2$ assumes the form

$$i\hbar \frac{\partial \psi(x,t)}{\partial t} = -\frac{\hbar^2}{2m}\frac{\partial^2 \psi(x,t)}{\partial x^2} + \frac{1}{2}m\omega^2 x^2 \psi(x,t) \tag{26}$$

and its quantum state $\psi(x,t)$ is found to be [36]

$$\psi(x,t) = \frac{\sqrt{a}}{\pi^{1/4}} \exp\left[-\frac{1}{2}a^2(x - x_0 \cos \omega t)^2 - i\frac{1}{2}\omega t - ia^2\left(xx_0 \sin \omega t - \frac{1}{4}x_0^2 \sin 2\omega t\right)\right] \tag{27}$$

where

$$a = \sqrt{\frac{m\omega}{\hbar}} \tag{28}$$

The amplitude square $|\psi(x,t)|^2$ of the wave-function given by equation (27) gives the position probability density

$$p(x,t) = |\psi(x,t)|^2 = \psi(x,t)\psi^*(x,t) = \frac{a}{\sqrt{\pi}} e^{-a^2(x - x_0 \cos \omega t)^2} \tag{29}$$

with a time-dependent mean-value, $\mu(t) = x_0 \cos \omega t$.

By setting $\sigma^2 = 1/(2a^2)$ so that

$$\frac{1}{2}\frac{1}{\sigma^2} = a^2 = \frac{m\omega}{\hbar} \tag{30}$$

equation (29) assumes the standard expression of a Gaussian distribution given by equation (1).

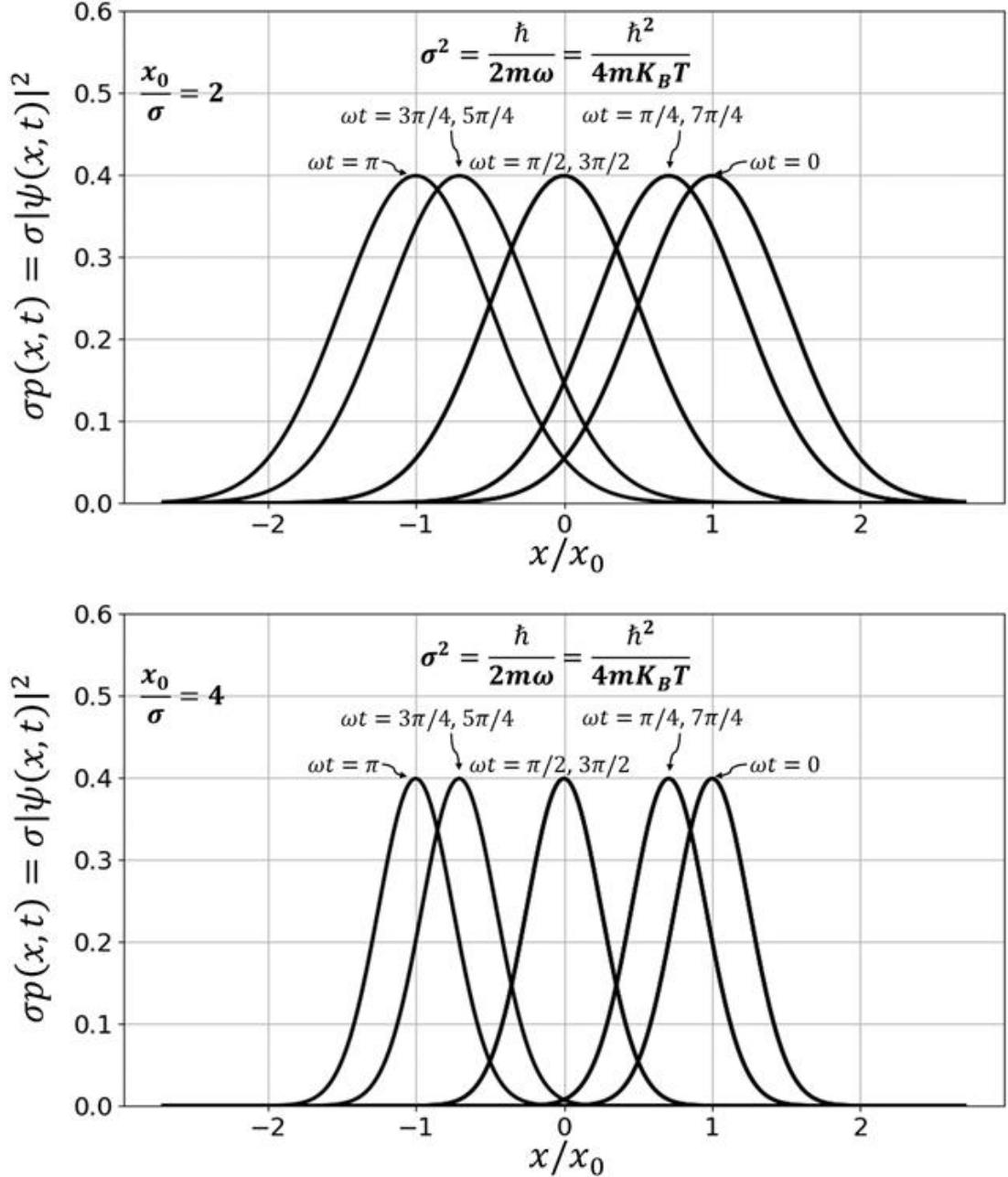

**Figure 5.** Evolution of the position probability density given by equation (31) of a quantum particle trapped in undamped harmonic potential, $V = \frac{1}{2}kx^2 = \frac{1}{2}m\omega^2 x^2$. Top: $\frac{x_0}{\sigma} = 2$. Bottom: Wider well with $\frac{x_0}{\sigma} = 4$. The position distribution density given by equation (31) has a constant variance $\sigma^2 = \hbar/(2m\omega) = \hbar^2/(4mK_BT)$ and oscillates perpetually within the undamped parabolic quantum well from $x = x_0$ to $x = -x_0$.

$$p(x,t) = |\psi(x,t)|^2 = \frac{1}{\sigma\sqrt{2\pi}} e^{-\frac{1}{2}\frac{x_0^2}{\sigma^2}\left(\frac{x}{x_0} - \cos\omega t\right)^2} \tag{31}$$

Returning to equation (11), and upon replacing $1/\sigma^2$ with $2m\omega/\hbar$ from equation (30), one obtains

$$\frac{\hbar^2}{8m\sigma^2} = \frac{\hbar\omega}{4} = \frac{1}{2}K_BT \tag{32}$$

Equation (32) yields that for a Brownian nanoparticle trapped in an undamped harmonic potential to satisfy the Schrödinger equation (26), its average kinetic energy $\frac{1}{2}K_BT = \frac{1}{2}\left(\frac{\hbar\omega}{2}\right)$, that is equal to ½ the energy of the ground state, $E_0 = \frac{1}{2}\hbar\omega$ of the quantum harmonic oscillator. This result is precisely the same as the long-term result offered by equation (23) that we reached with a top down route upon integrating the Fokker-Planck equation (18). At the limiting case when the viscosity $\eta$ tends to zero (undamped oscillator), equation (23) yields the result of equation (32) that was obtained with a bottom-up route where we calculated the position probability density of the quantum wave pocket given by equation (29) or (31).

Equation (32) in association with the quadratic dispersion relation that results from the Hamiltonian of the quantum/Brownian particle [36, 37]

$$\omega = \frac{\hbar}{2m}\left(\frac{2\pi}{\lambda}\right)^2 \tag{33}$$

yields directly the de Broglie wave-length of the quantum/Brownian nanoparticle given by equation (24) and plotted in Figure 3. Substitution of the expression of the frequency, $\omega = 2K_BT/\hbar$ given equation (32), which is now derived with the bottom-up route, into the expression of the shear modulus $G = \frac{k}{6\pi R} = \frac{m\omega^2}{6\pi R}$, yields again the expression offered by equation (25) and plotted in Figure 4.

Figure 5 plots the evolution with time of the position distribution density function $p(x,t) = |\psi(x,t)|^2$ given by equation (31) at times $\omega t = 0$, π/4, π/2, 3π/4, π, 5π/4, 3π/2 and 7π/4. The position distribution density function has a constant variance $\sigma^2 = \hbar/(2m\omega) = \hbar^2/(4mK_BT)$, which oscillates perpetually within the parabolic quantum well from $x = x_0$ to $x = -x_0$.

**CONCLUSIONS**

In view that de Broglie matter waves have been experimentally observed for molecules as large as few nanometers [30-34], in this paper we examined whether Brownian particles with comparable size as small as few nanometers that undergo continuous thermally driven fluctuations may manifest a particle-wave duality. We first show that a necessary condition for a Brownian particle with mass, $m$ to manifest a particle-wave duality is that the variance $\sigma^2$ of its position probability density shall be a constant, independent of time - that is $\sigma^2 = \hbar^2/(4mK_BT)$. Accordingly, Brownian particles immersed in a Newtonian viscous fluid, where their variance $\sigma^2(t) = 2NDt$, grows linearly with time cannot manifest a particle-wave duality regardless how small their size is. This finding agrees with past conclusions by Grebert et al. [38], that quantum mechanics is not equivalent to a Markovian diffusion process. In contrast, the paper shows that when a Brownian particle is embedded in an elastic solid, trapped in a harmonic potential well, with or without dissipation, the requirement for the nanoparticle to manifest a particle-wave duality leads to the physically acceptable result that its mean kinetic energy, $\frac{1}{2}K_BT = \frac{1}{4}\hbar\omega = \frac{1}{2}E_0$, where $E_0 = \frac{1}{2}\hbar\omega$ is the ground-state energy of the quantum harmonic oscillator. This result is reached by following two independent routes. First, we follow a top-down approach where the position probability density, $p(x,t)$ of a Brownian particle trapped in a harmonic potential with dissipation is described with the Fokker-Planck equation and the resulting Gaussian distribution function $p(x,t)$ is set equal to the position probability density from a matter-wave $|\psi(x,t)|^2 = \psi_1^2(x,t) + \psi_2^2(x,t)$. Second, we follow a bottom-up approach by examining the solution of an oscillating quantum particle (wave pocket) trapped in a harmonic potential [36, 37] and by demanding that the average kinetic energy of the oscillating wave-pocket to be equal to the average kinetic energy of a Brownian nanoparticle $=\frac{1}{2}K_BT$. Our calculations from both routes reach precisely the same results and show that for this to happen the trapping needs to be very strong, so that a Brownian nanoparticle with mass $m$ and radius $R$ needs to be embedded in an extremely stiff solid with shear modulus, $G$ proportional to $(m/R)(K_BT/\hbar)^2$. Cooling reduces the value of the shear modulus $G$ needed to offer the required trapping; nevertheless, in view of the various challenges associated with cooling schemes for interferometry with massive particles [33, 34, 53, 54], our study concludes that the potential observation of de Broglie matter-waves emitted from trapped Brownian nanoparticles remains a formidable task.

14. Einstein, A., 1905. Über die von der molekularkinetischen Theorie der Wärme geforderte Bewegung von in ruhenden Flüssigkeiten suspendierten Teilchen. *Annalen der Physik, 4*.

15. Pathria R. K. (1996) *Statistical Mechanics,* Second edn, Butterworth Heinemann, Oxford, UK.

16. Risken, H., 1996. Fokker-Planck equation for several variables; methods of solution. In *The Fokker-Planck Equation: Methods of Solution and Applications* (pp. 133-162). Berlin, Heidelberg: Springer Berlin Heidelberg.

17. Araujo, M.T. and Drigo Filho, E., 2012. A general solution of the Fokker-Planck equation. *Journal of Statistical Physics, 146*, pp.610-619.

18. Polotto, F., Drigo Filho, E., Chahine, J. and de Oliveira, R.J., 2018. Supersymmetric quantum mechanics method for the Fokker–Planck equation with applications to protein folding dynamics. *Physica A: Statistical Mechanics and its Applications, 493*, pp.286-300.

19. Santra, I., Das, S. and Nath, S.K., 2021. Brownian motion under intermittent harmonic potentials. *Journal of Physics A: Mathematical and Theoretical, 54*(33), p.334001.

20. Langevin, P., 1908. Sur la théorie du mouvement brownien. *Compt. Rendus, 146,* pp.530-533.

21. Uhlenbeck, G.E. and Ornstein, L.S., 1930. On the theory of the Brownian motion. *Physical Review, 36*(5), p.823.

22. Wang, M.C. and Uhlenbeck, G.E., 1945. On the theory of the Brownian motion II. *Reviews of Modern Physics, 17*(2-3), p.323.

23. Landau L.D. and Lifshitz E.m. (1980) Chapter V in *Statistical Physics* Part 1 Vol. 5 of Course of Theoretical Physics, Second Revised and Enlarged Edition, Pergamon Press, Oxford, UK.

24. Attard, P., 2012. *Non-equilibrium thermodynamics and statistical mechanics: Foundations and applications.* OUP Oxford.

25. Coffey, W. and Kalmykov, Y.P., 2012. *The Langevin equation: with applications to stochastic problems in physics, chemistry and electrical engineering* (Vol. 27). World Scientific.

26. de Broglie, L., 1925. Research on the theory of quanta. In *Annales de Physique* (Vol. 10, No. 3, pp. 22-128).
19